\documentstyle[aasms4,12pt]{article}
\begin{document}

\title{The Importance of Realistic Starting Models for Hydrodynamic Simulations
of Stellar Collisions}

\author{Alison Sills} \affil{Department of Astronomy, Yale University,
P.O. Box 208101, New Haven, CT, 06520-8101, USA}

\author{James C. Lombardi, Jr.\altaffilmark{1}} \affil{Center for
Radiophysics and Space Research, Cornell University, Ithaca, NY 14853,
USA}

\altaffiltext{1}{also Department of Astronomy, Cornell University}
 
\begin{abstract}
We demonstrate the necessity of using realistic stellar models taken
from stellar evolution codes, as opposed to polytropes, for starting
models in smoothed particle hydrodynamics calculations of collisions
between main sequence stars. Evolved stars have mean molecular weight
gradients, which affect their entropy profiles and therefore affect
how they react during a collision. The structure of stellar collision
products of polytrope parent stars is significantly different from
that of collision products of realistic parent models. These
differences strongly affect the future evolution of the collision
products, particularly products of collisions between unequal mass
stars which have undergone significant chemical evolution. The use of
polytropes as parent star models is likely to result in qualitatively
mistaken results for the structure of the collision product.
\end{abstract}


\section{Introduction}

Hydrodynamical codes provide a useful numerical tool for determining
what happens during a collision between two main sequence stars. In
particular, the smoothed particle hydrodynamic (SPH) technique has
been used extensively in the past decade to investigate the
possibility that blue stragglers can be created by stellar collisions
in globular clusters (Benz \& Hills 1987, Goodmann \& Hernquist 1991,
Lombardi, Rasio \& Shapiro 1996, Sandquist, Bolte \& Hernquist
1997). The emphasis of these groups has been to determine the internal
structure of stellar collision products. These models can then be
compared to blue stragglers in globular clusters, to help determine
the creation mechanisms for blue stragglers, or to give information
about the dynamics of the cluster as a whole.

When running an SPH simulation of a stellar collision, one needs to
specify the entropy (or pressure and density) profiles of the parent
stars. Most groups used polytropes with polytropic indices of $n=1.5$
(for convective stars) and $n=3$ (for radiative stars) as
approximations to the stars. While these polytropes are reasonable
approximations to zero age main sequence stars, they are not
sufficient representations of evolved stars. The most important
difference between the two kinds of stellar models is that evolved
stars have gradients of mean molecular weight. In this paper, we deal
specifically with main sequence parent stars, and not with white
dwarfs which are well represented by $n=3$ polytropes.

An SPH code takes the two parent stars and divides them up into many
fluid elements.  For head on collisions, the code creates a collision
product by, in essence, sorting the fluid in a configuration with the
quantity $A=P/\rho^{\Gamma_1}$ increasing outward (the dynamical
stability criterion), where $P$ is pressure, $\rho$ is density and the
adiabatic exponent $\Gamma_1=5/3$ for an ideal gas.  While real stars
are not exactly ideal gases, the low mass stars we are considering
here are dominated by gas pressure, with radiation pressure
contributing at most 5\% to the total pressure.  The quantity $A$ is
closely related to entropy, since in the absence of shocks both are
conserved by fluid elements during a collision.  For convenience, we
will refer to $A$ itself as the entropy.  The entropy of a fluid
element can be increased by shock heating, but since collisions in
globular clusters are relatively gentle, this effect is not dominant.
Therefore, sorting the two parent stars by entropy yields a reasonable
guess for the product of a given head on collision.

During the main sequence lifetime of a star, its central temperature
remains almost constant, since nuclear burning is a self-regulating
process. The central pressure of a star is also approximately
constant, since this is the pressure which is necessary to support a
star of that mass. Therefore, for a star with an ideal gas equation of
state $P=\frac{R \rho T}{\mu}$, the ratio of density to mean molecular
weight must also remain constant. As the mean molecular weight
increases, the density must also increase in order to keep the
pressure and temperature constant. Since the density is in the
denominator of the entropy, and since it is raised to a power greater
than one, any increase in density results in a significant decrease in
entropy.  An evolved star has burnt much of its hydrogen to helium,
which increases the mean molecular weight of the central material of
the star. The central entropy of this evolved star will be
significantly less than that of a corresponding unevolved star, or a
polytrope.

\section{A Demonstration}

Take, for example, a $0.8 M_{\odot}$ star at an age of 15 Gyr, a
turnoff star in a typical globular cluster. It has burnt almost all of
its central hydrogen to helium, increasing the mean molecular weight
at the center by a factor of about 2.5. Since the mean molecular
weight at the center has increased, the central entropy has decreased
(see Figure 1). The realistic stellar models have been calculated
using the Yale Rotating Evolution Code (YREC, Guenther {\it et al.}
1992). The polytrope used to model an unevolved $0.8 M_{\odot}$ star
has a polytropic index of $n=3$.

Also consider a $0.4 M_{\odot}$ star at an age of 15 Gyr. This star
has completed only a small fraction of its main sequence lifetime, and
therefore has not changed its mean molecular weight profile
significantly. The evolved entropy profile looks quite similar to the
entropy profile of the $n=1.5$ polytrope approximation of an unevolved
star of the same mass (see Figure 1). The decrease in entropy near the
surface is a result of approximating the equation of state as that of
a fully ionized ideal gas.

From figure 1, we can ascertain the general properties of a product
that would result from the collision between any two of these
stars. The portions of the two stars with the lowest entropy would
settle to the center of the collision product, while that with the
highest entropy would end at the surface. Therefore, if we use the two
polytrope models as our parent stars, the entire $0.4 M_{\odot}$ star
will become the new core of the collision product, displacing the $0.8
M_{\odot}$ star outward. The collision product would then have a core
of unburnt hydrogen from the $0.4 M_{\odot}$ star, surrounded by a
layer of helium-rich material from the core of the $0.8 M_{\odot}$
star.  However, a significantly different outcome occurs if we use the
evolved parent stars. In this case, the center of the $0.8 M_{\odot}$
star has the lowest entropy and consequentially sinks to the center of the
collision product.  The core of the $0.4 M_{\odot}$ star has a much
higher entropy, and ends up further out in the collision product. This
star will have a core which is depleted of hydrogen from the core of
the $0.8 M_{\odot}$ star, and an outer layer which is almost
completely unburnt.

We can also consider a collision between two $0.8 M_{\odot}$ stars. We
expect in this case that there will be no significant differences in
the fluid distribution between using polytropes and using realistic
parent stars. Although the values of the entropy curves are different,
the overall shape is generally the same: the lowest entropy occurs in
the center of the star, and the entropy increases monotonically to the
surface. Therefore, when two of these stars collide, the centers of
the two stars will fall to the center of the product, and the outer
layers will stay on the outside.

We have performed new simulations of a head-on collision between a
$0.8 M_{\odot}$ and a $0.4 M_{\odot}$ star, and between two $0.8
M_{\odot}$ stars using a modified version of the SPH code developed by
Rasio (1991). The parent models were created using the density and
pressure profiles from the realistic evolved models calculated by
YREC.  The $0.4 M_{\odot}$ and $0.8 M_{\odot}$ stars were represented
by $7.5 \times 10^3$ and $1.5 \times 10^4$ equal mass SPH particles
respectively. Details of the SPH scheme are presented in Lombardi {\it
et al.} (1996).  We compared the results of these calculations to the
results of previous SPH calculations which used polytropes as parent
stars (see cases A and G in Lombardi, Rasio \& Shapiro 1996). The
structures of the merger products are compared in figure 2 (for the
$0.8 M_{\odot} + 0.4 M_{\odot}$ collision) and figure 3 ($0.8
M_{\odot} + 0.8 M_{\odot}$). The results are essentially as we
predicted above. This can be seen best in the plots of helium fraction
$Y$ versus mass fraction, since the highest $Y$ material was in the
center of the $0.8 M_{\odot}$ star at the beginning of the collision.

Using the technique outlined in Sills {\it et al.} 1997, we have used
the results of these collisions as the starting models for stellar
evolution calculations.  The results are presented in figure 4. As
expected, the evolutionary tracks for the two cases of the $0.8
M_{\odot} + 0.4 M_{\odot}$ collision are very different. The collision
of two polytropes contains a significant amount of hydrogen in its
core (from the $0.4 M_{\odot}$ parent). Therefore, the evolutionary
track of this collision product shows a significant main sequence of
$1.4\times 10^9$ years from point 1 to point 2. However, the product
of the collision between realistic parent stars did not have much
hydrogen in its core, and so its main sequence (point 1 to point 2) is
almost non-existent, lasting only $3.4\times 10^6$ years. The two
tracks for the $0.8 M_{\odot} + 0.8 M_{\odot}$ collisions are
essentially the same from the main sequence onward since the chemical
composition and temperature profiles of the collision product were not
significantly affected by the choice of parent star.

\section{Conclusions}

We advocate the use of realistic stellar models, taken from stellar
evolution codes, for starting models in hydrodynamic calculations.  In
particular, polytropic models are insufficient to model main sequence
and giant stars which have undergone significant chemical evolution.
While previous groups have made reasonable guesses for the structure
of turnoff mass stars in globular clusters, the fact that they did not
use realistic parent models significantly affects the structure of the
collision product. Therefore, any further evolution of these stars
does not adequately represent the evolution of a real collision
between globular cluster stars. The differences will be significant
for work which draws conclusions about the structure of the collision
product or which uses the results of hydrodynamic calculations as the
starting models for stellar evolution calculations.

The general conclusions of Lombardi {\it et al.} 1996 and Sills {\it et
al.} 1997 are not invalidated by this work. It is still true that
significant mixing does not occur either during the collision or
during the thermal relaxation phase after the collision. However, the
lifetime of the star on the main sequence and its subsequent evolution
is significantly different from previous work, especially in
collisions between unequal mass parent stars.

\acknowledgements This work was supported by NASA grant NAG5-2867 and
NSF grant AST-9357387 at Yale University, as well as NSF grant AST
91-19475 and NASA grant NAG5-2809 at Cornell University.  Hydrodynamic
simulations were performed at the Cornell Theory Center, which
receives major funding from the NSF and IBM Corporation, with
additional support from the New York State Science and Technology
Foundation and members of the Corporate Research Institute.  We would
like to thank Fred Rasio and Charles Bailyn for their helpful
discussions.

\clearpage 

\figcaption[entropy.eps]{Entropy profiles of polytropes and realistic
parent stars evolved to 15 Gyr. Two masses are shown here: $0.4
M_{\odot}$ (dashed and dot-dashed lines) and $0.8 M_{\odot}$ (solid
and dotted lines). Note the difference in shape between the polytrope
and realistic curves of the same mass. The pressure $P$ and density
$\rho$ are evaluated in cgs units.}

\figcaption[Gstructure.eps]{Comparison of the structure of the
collision products for an $0.8 M_{\odot}+0.4 M_{\odot}$
collision. Solid line is the product of two realistic parent stars,
and the dotted line is the product of two polytrope models. Note
especially the large difference in chemical profiles between the two
cases.}

\figcaption[Astructure.eps]{Same as figure 2 for a $0.8 M_{\odot}+0.8
M_{\odot}$ collision. Since these stars are the same mass, there are
only subtle differences between the polytrope and the realistic parent
star collisions.}

\figcaption[tracks.eps]{Evolutionary tracks of the merger products
presented in figures 2 and 3. Solid lines are the tracks for collision
products from realistic models, and dotted lines are from
polytropes. The upper panel shows tracks for collisions between a $0.8
M_{\odot}$ and a $0.4 M_{\odot}$ star, while the lower panel shows
tracks for a collision between two $0.8 M_{\odot}$ stars.  The labels
show significant evolutionary points along the track. Point 0 is the
position of the collision product immediately after the
collision. Point 1 is the equivalent of the zero age main sequence,
and point 2 is the terminal age main sequence, when the star has no
more hydrogen in its core. Point 3 is the base of the giant branch. The
main sequence lifetimes (between points 1 and 2) of each star are given.}

\clearpage
\plotone{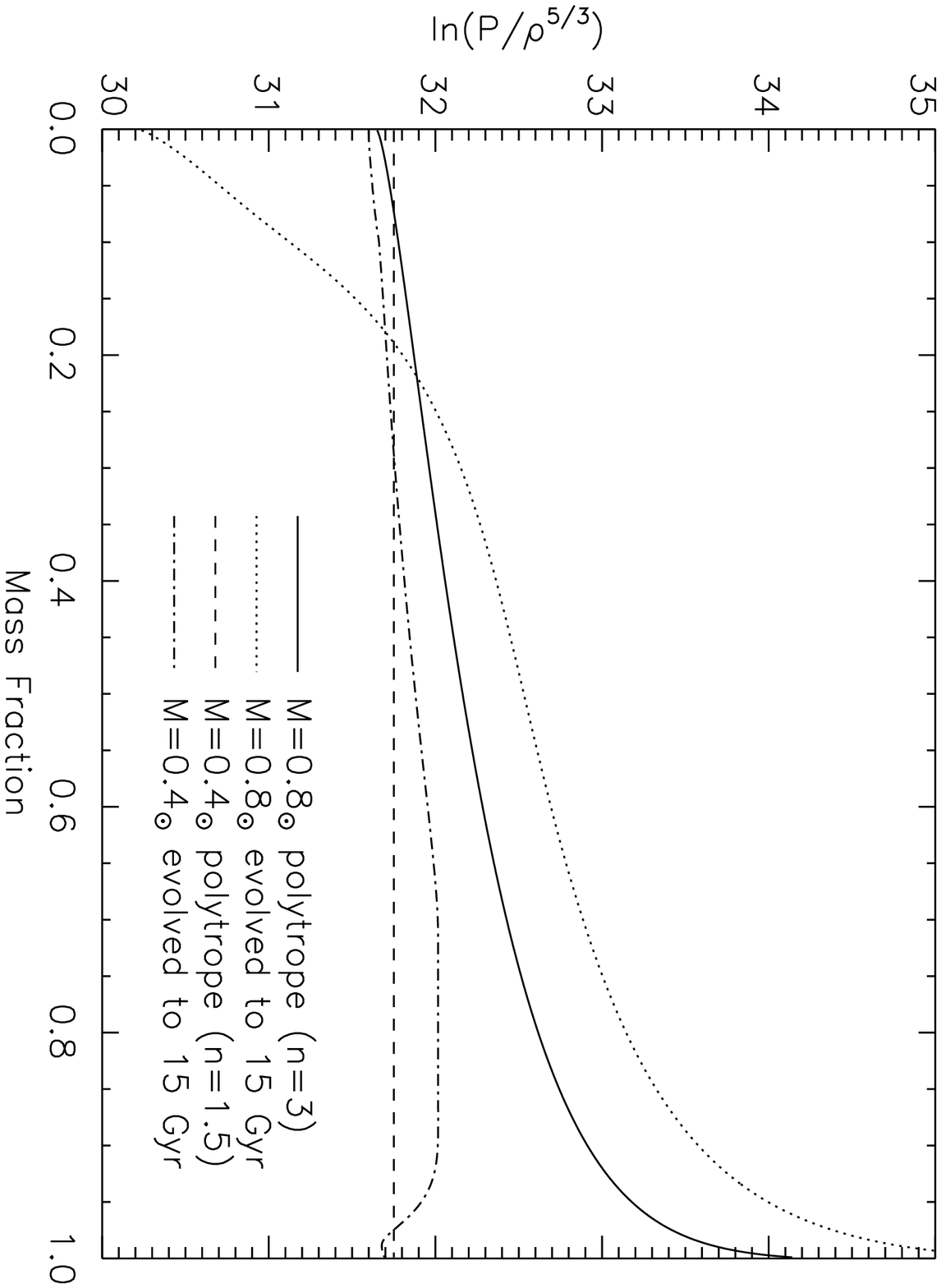}
\clearpage
\plotone{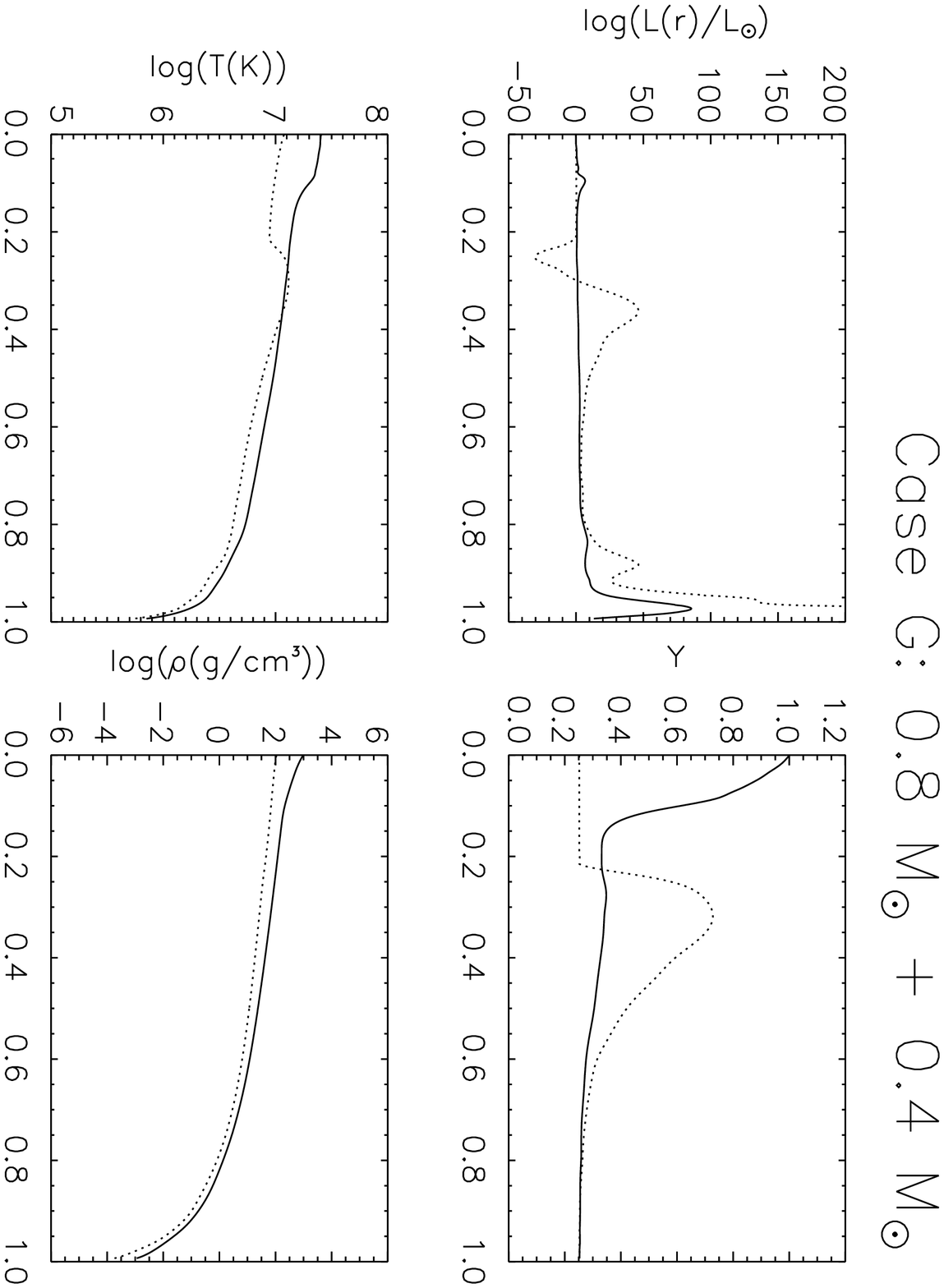}
\clearpage
\plotone{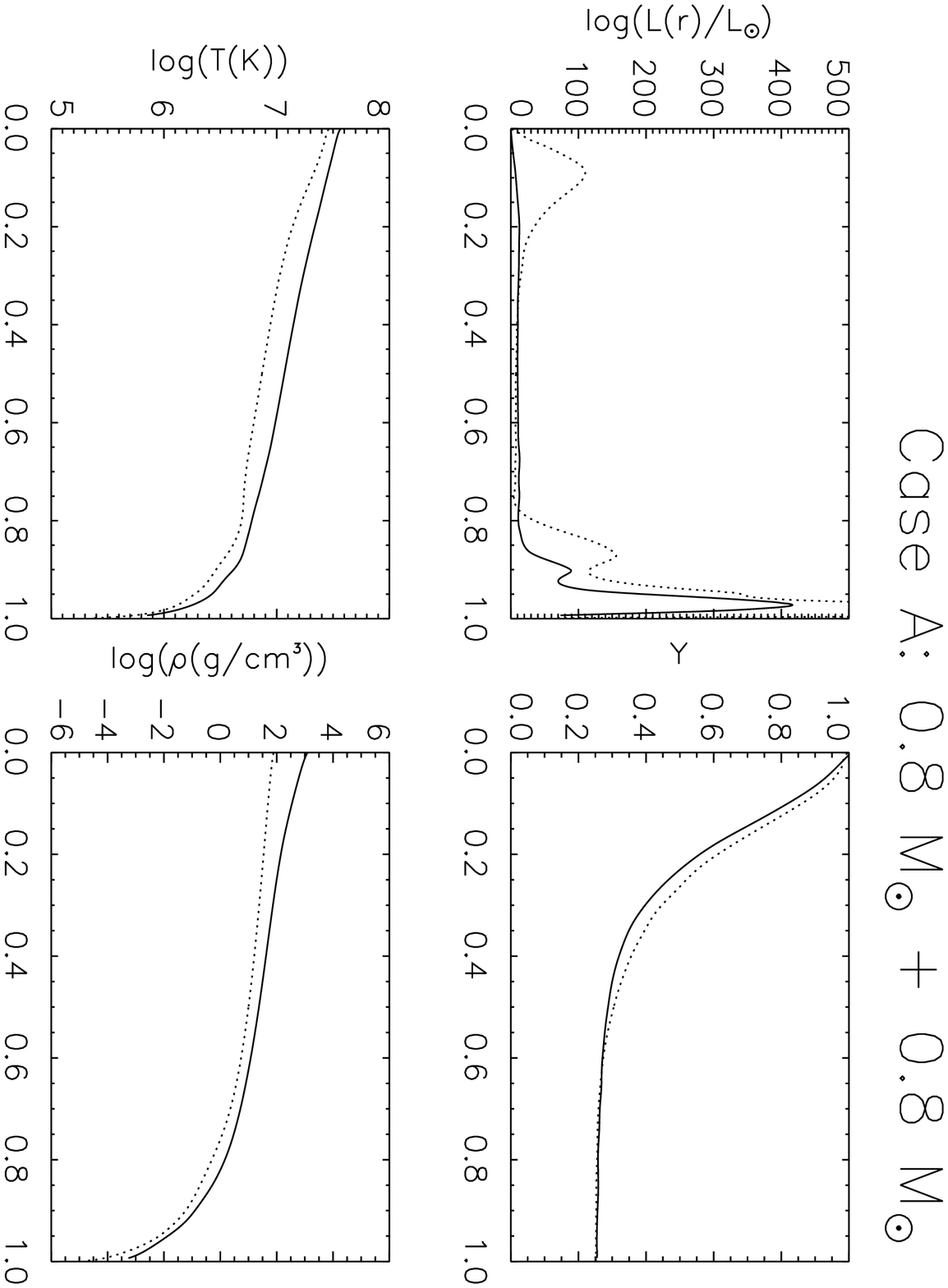}
\clearpage
\plotone{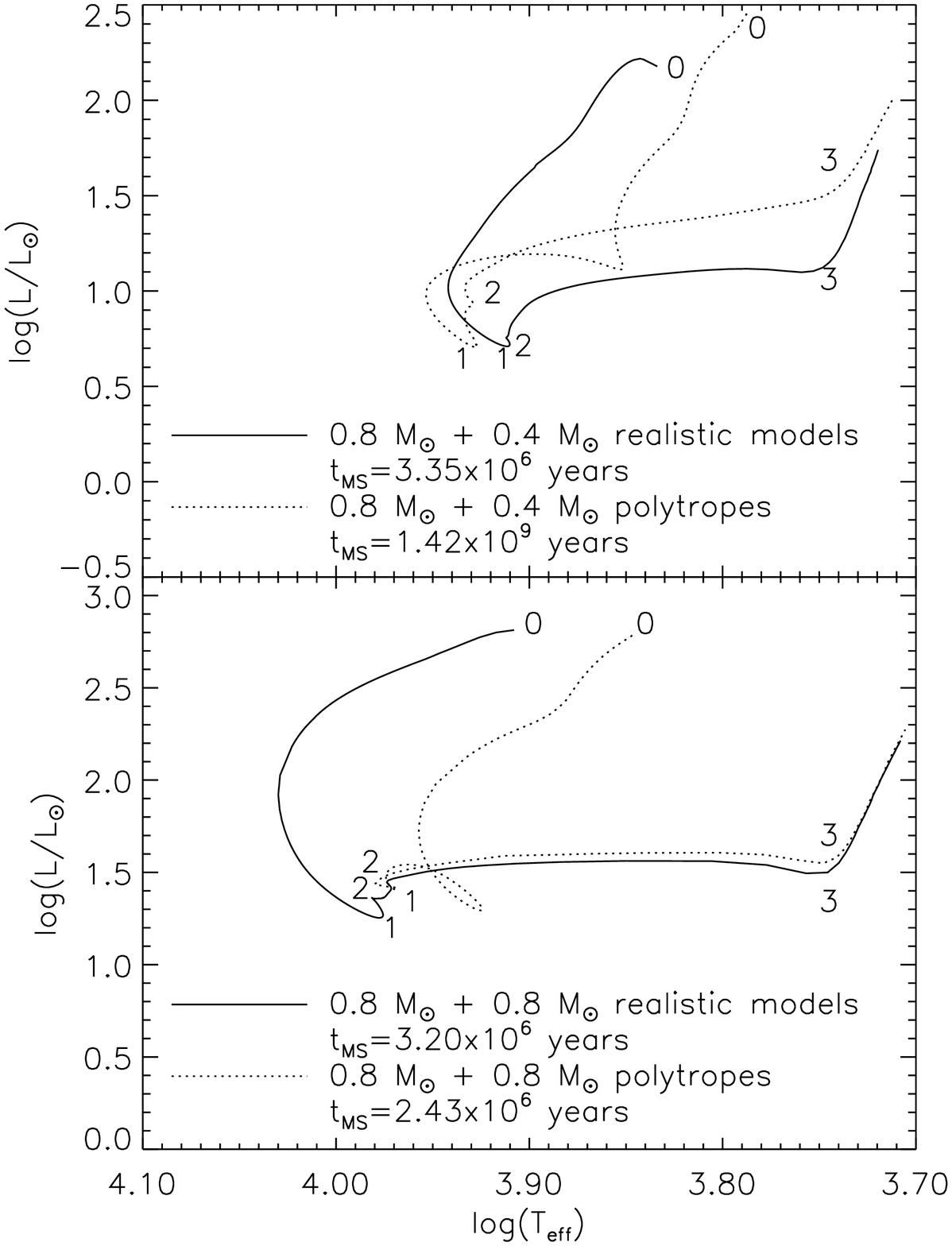}

\end{document}